\documentclass[onecolumn,tightenlines,showpacs,nofootinbib,preprintnumbers,
amsmath,amssymb]{revtex4}

\usepackage[usenames,dvipsnames]{color}
\usepackage{caption2}
\usepackage{dcolumn}
\usepackage{graphicx}
\usepackage{subfigure}
\usepackage{epstopdf}
\usepackage{bm}
\usepackage{lscape}

\usepackage{setspace}
\begin{document}
\begin{spacing}{1.5}

\title{Direct $CP$ violation from isospin symmetry breaking effects in PQCD}

\author{Gang L\"{u}$^{1}$\footnote{Email: ganglv66@sina.com}, Qin-Qin Zhi $^{1}$\footnote{Email: zhiqinqin11@163.com}}

\affiliation{\small $^{1}$College of Science, Henan University of Technology, Zhengzhou 450001, China\\
}

\begin{abstract}
We investigate the direct $CP$ violation for the decay process of
$\bar{B}_{s}\rightarrow P(V)\pi^{0}$ (P,V refer to the pseudoscalar meson and vector meson, respectively)
via isospin symmetry breaking effects from the $\pi^{0}-\eta-\eta'$ mixing mechanism in PQCD factorization approach.
Isospin symmetry breaking arises from the electroweak interaction and the u-d quark mass difference by the strong interaction
which are known to be tiny. However, we find that isospin symmetry breaking at the leading order shifts
the $CP$ violation due to the new strong phases.
\end{abstract}

\maketitle

\section{\label{intro}Introduction}
The measurement of $CP$ violation is an important area in understanding Standard Model (SM)
and  exploring new physics signals. Cabibbo-Kobayashi-Maskawa (CKM) matrix \cite{cab,kob}
due to the quark flavour mixing provides us the weak phases. The weak phase associated with the strong phase
is responsible for the source of the CP violation. The strong phase comes from the dynamics of QCD and
the other mechanism.

The hadronic matrix elements of the nonleptonic weak decay are known to be associated with the
strong phase. We can estimate the power contribution by the factorization method in the limit of
$1/m_{b}$ ($m_{b}$ refers to b quark mass) in B meson decay process.
Basing on the QCD correction and taking into account transverse momenta,
PQCD factorization method safely avoids the infrared divergence by introducing the Sudakov factor
which is applied to deal with the decay amplitude related with the hadronic matrix elements.
The decay amplitude can be written as the convolution of the meson wave functions and
the hard kernel, which show the contributions of the non perturbative and the perturbative parts, respectively \cite{AC,AG,YA,LKM}.

Isospin symmetry plays an important part in the weak decay process of B meson. We can infer sum rule associated with
the isospin symmetry to form a triangular shape on a complex plane for the decay amplitude.
One can eliminate uncertainty from the penguin diagram by the isospin analysis in B decays \cite{MGronau1990}.
Isospin symmetry breaking via $\rho$-$\omega$ mixing produces the strong phase to lead to the large CP violation
in the three bodies decay process \cite{gar1998,gang4}. Isospin symmetry is approximate symmetry due to identical
u and d quark masses in Standard Model (SM). The mixing of pseudoscalar mesons $\pi^{0}$-$\eta$-$\eta'$  is  from the isospin symmetry breaking within QCD.
Isospin symmetry breaking plays a significant role for the decays of $B\rightarrow \pi\pi$, which
breaks the triangle relationship in the framework of generalized factorization \cite{garden1999M}.
$\pi^{0}$-$\eta$-$\eta'$ mixing is discussed by the model-independent way in $B\rightarrow \pi\pi$ decay process
using flavor SU(3) symmetry \cite{Gronau2005J}. The quark-flavor mixing produces the
$\pi^{0}$-$\eta$-$\eta'$ mixing due to the isospin symmetry breaking \cite{Kroll2005PP}.
Recently, isospin symmetry breaking is discussed by incorporating the Nambu-Jona-Lasinio model
in a generalized multiquark interaction scheme \cite{AOsip2016}.
However, one can find that the research
rarely pays attention to the CP violation
from the effect of isospin symmetry breaking via $\pi^{0}$-$\eta$-$\eta'$ mixing.
The strong phase may be introduced to affect the value of CP violation accordingly
which is similar with the contribution from the isospin symmetry breaking by
the $\rho$-$\omega$ mixing \cite{gar1998,gang4}.

The remainder of this paper is organized as follows. In Sec.
\ref{sec:hamckm} we present the form of the effective Hamiltonian.
In Sec. \ref{sec:cpv1} we give the calculating formalism of $CP$ violation from isospin symmetry breaking
in $\bar{B}_{s}\rightarrow P(v)\pi^0$.
Input parameters are presented in Sec.\ref{input}.
We present the numerical results in Sec.\ref{num}.
Summary and discussion are included in
Sec. \ref{sum}. The related functions defined in the text are given
in the Appendix.

\section{\label{sec:hamckm}The effective hamiltonian}

With the operator product expansion, the effective weak Hamiltonian can
be written as \cite{buras}
\begin{eqnarray}
{\cal H}_{\Delta B=1} = {G_F\over \sqrt{2}}[
V_{ub}V^*_{ud}(C_1 O^u_1 + C_2 O^u_2)    \nonumber   \\
 - V_{tb}V^*_{td}\sum^{10}_{i=3} C_i O_i] + H.C.,\;
\label{2a}
\vspace{2mm}
\end{eqnarray}
where $G_F$ represents Fermi constant, $C_i$ (i=1,...,10) are the Wilson coefficients, $V_{ub}$,
$V_{ud}$, $V_{tb}$ and $V_{td}$ are the CKM matrix elements. The
operators $O_i$ have the following forms:
\begin{eqnarray}
\begin{split}
O^{u}_1&= \bar d_\alpha \gamma_\mu(1-\gamma_5)u_\beta\bar
u_\beta\gamma^\mu(1-\gamma_5)b_\alpha,  \\
O^{u}_2&= \bar d \gamma_\mu(1-\gamma_5)u\bar
u\gamma^\mu(1-\gamma_5)b,  \\
O_3&= \bar d \gamma_\mu(1-\gamma_5)b \sum_{q'}
\bar q' \gamma^\mu(1-\gamma_5) q',  \\
O_4 &= \bar d_\alpha \gamma_\mu(1-\gamma_5)b_\beta \sum_{q'}
\bar q'_\beta \gamma^\mu(1-\gamma_5) q'_\alpha,  \\
O_5&= \bar d \gamma_\mu(1-\gamma_5)b \sum_{q'} \bar q'
\gamma^\mu(1+\gamma_5)q',  \\
O_6& = \bar d_\alpha \gamma_\mu(1-\gamma_5)b_\beta \sum_{q'}
\bar q'_\beta \gamma^\mu(1+\gamma_5) q'_\alpha,  \\
O_7&= \frac{3}{2}\bar d \gamma_\mu(1-\gamma_5)b \sum_{q'}
e_{q'}\bar q' \gamma^\mu(1+\gamma_5) q',  \\
O_8 &= \frac{3}{2} \bar d_\alpha \gamma_\mu(1-\gamma_5)b_\beta \sum_{q'}
e_{q'}\bar q'_\beta \gamma^\mu(1+\gamma_5) q'_\alpha,  \\
O_9&= \frac{3}{2}\bar d \gamma_\mu(1-\gamma_5)b \sum_{q'} e_{q'}\bar q'
\gamma^\mu(1-\gamma_5)q',  \\
O_{10}& = \frac{3}{2}\bar d_\alpha \gamma_\mu(1-\gamma_5)b_\beta \sum_{q'}
e_{q'}\bar q'_\beta \gamma^\mu(1-\gamma_5) q'_\alpha.
\label{2b}
\end{split}
\end{eqnarray}
where $\alpha$ and $\beta$ are color indices, and $q^\prime=u, d$
or $s$ quarks. In Eq.(\ref{2b}) $O_1^u$ and $O_2^u$ are tree
operators, $O_3$--$O_6$ are QCD penguin operators and $O_7$--$O_{10}$ are
the operators associated with electroweak penguin diagrams.

We can obtain numerical values of $C_i$. When $C_i(m_b)$ \cite{LKM},
\begin{eqnarray}
\begin{split}
C_1 &=-0.2703, \;\; \;C_2=1.1188,  \\
C_3 &= 0.0126,\;\;\;C_4 = -0.0270,  \\
C_5 &= 0.0085,\;\;\;C_6 = -0.0326,  \\
C_7 &= 0.0011,\;\;\;C_8 = 0.0004,  \\
C_9&= -0.0090,\;\;\;C_{10} = 0.0022.
\label{2k}
\end{split}
\end{eqnarray}

One can obtain numerical values of $a_i$. The combinations $a_i$ of Wilson coefficients are defined as \cite{AG, YH}
\begin{eqnarray}
\begin{split}
a_1 &= C_2+C_1/ 3, \;\; \;a_2 = C_1+C_2/ 3,  \\
a_3 &= C_3+C_4/ 3,\;\;\;a_4 = C_4+C_3/ 3,  \\
a_5 &= C_5+C_6/ 3,\;\;\;a_6 = C_6+C_5/ 3,  \\
a_7 &= C_7+C_8/ 3,\;\;\;a_8 = C_8+C_7/ 3,  \\
a_9&= C_9+C_{10}/ 3,\;\;\;a_{10} = C_{10}+C_9/ 3.  \\
\label{2k}
\end{split}
\end{eqnarray}

\section{\label{sec:cpv1}$CP$ violation from Isospin symmetry breaking effects}
\subsection{\label{subsec:form}Formalism}

It is convenient to introduce isospin vector triplet $\pi_{3}$, isospin scalar $\eta_{n}$
and isospin scalar $\eta_{s}$ which can be distinguished by including strange quark or not.
The $SU(3)$ singlet $\eta_{0}$ and octet $\eta_{8}$ can be well described by the translation
$\eta_{n}=\frac{\sqrt{2}\eta_{0}+\eta_{8}}{\sqrt{3}}$ and $\eta_{s}=\sqrt{\frac{1}{3}}\eta_{0}-\sqrt{\frac{2}{3}}\eta_{8}$.
The states of $\pi_{3}$, $\eta_{n}$ and $\eta_{s}$ are identified by $\pi_{3}=\frac{1}{\sqrt{2}}|u\bar{u}-d\bar{d}>$, $\eta_{n}=\frac{1}{\sqrt{2}}|u\bar{u}+d\bar{d}>$
and $\eta_{s}=|s\bar{s}>$ which are obtained from the quark model, respectively.
The physical meson states can be transformed from the $\pi_{3}$, $\eta_{n}$ and $\eta_{s}$ by unitary matrix $U$ \cite{ Kroll2005PP}:
\begin{gather}
\label{Ubian}
\begin{pmatrix} \pi^{0}\\ \eta \\ \eta'\end{pmatrix}=U(\varepsilon_{1},\varepsilon_{2},\phi)\begin{pmatrix} \pi_{3}\\ \eta_{n}\\ \eta_{s} \end{pmatrix},
\end{gather}
where
\begin{gather}
U(\varepsilon_{1},\varepsilon_{2},\phi)=\begin{pmatrix} 1 & \varepsilon_{1}+\varepsilon_{2}cos\phi & -\varepsilon_{2}sin\phi \\
-\varepsilon_{2}-\varepsilon_{1}cos\phi & cos\phi & -sin\phi \\ -\varepsilon_{1}sin\phi & sin\phi & cos\phi \end{pmatrix},
\label{pm}
\end{gather}
$\varepsilon_{1}$, $\varepsilon_{2}$$\propto O(\lambda),\lambda \ll 1$ and the higher order terms are neglected.
In the isospin limit of $\varepsilon_{1}\rightarrow 0,\varepsilon_{2}\rightarrow0$, we can find that the formula
is expressed as the $\eta-\eta'$ mixing in Eq.(\ref{1z}):
\begin{equation}
\left(\begin{matrix} \eta \\ \eta'
\end{matrix}
\right)=U'(\phi)\left(\begin{matrix} \eta_{n} \\ \eta_{s}
\end{matrix}
\right)
=
\left( \begin{array}{ccc}
\cos\phi & -sin\phi \\
\sin\phi  & cos\phi
\end{array} \right)
\left(\begin{matrix} \eta_{n} \\ \eta_{s}
\end{matrix}
\right),
\label{1z}
\end{equation}
where $\phi$ is the mixing angle \cite{AMLi2007}.
The $\eta$ and $\eta'$ mixing depends on the quark flavor basises $\eta_{n}$ and $\eta_{s}$.

The relevant decay constants can be written as \cite{tpbm}:
\begin{eqnarray}
\begin{split}
\langle 0|\bar n\gamma^\mu\gamma_5 n|\eta_n(P)\rangle&= \frac{i}{\sqrt2}\,f_n\,P^\mu \;, \\
\langle 0|\bar s\gamma^\mu\gamma_5 s|\eta_s(P)\rangle &= i f_s\,P^\mu\;,
\end{split}
\label{deffq}
\end{eqnarray}
where $P$ refers to the momenta of $\eta_n$ or $\eta_s$.

One can understand that isospin symmetry breaking comes from the electroweak interaction and $u-d$ quark mass difference in Stand Model.
We can calculate the isospin symmetry breaking correction by chiral perturbative theory which induces the $\pi^{0}-\eta-\eta'$ mixing.
To the leading order of isospin symmetry breaking, the physical eigenstate $\pi^{0}$, $\eta$ and $\eta'$ from Eq.(\ref{Ubian})(\ref{pm}) can be written as
\begin{eqnarray}
\begin{split}
|\pi^{0}\rangle&=|\pi_{3}\rangle+(\varepsilon_{1}+\varepsilon_{2}\cos\phi)|\eta_{n}\rangle-\varepsilon_{2}\sin\phi|\eta_{s}\rangle,\\
|\eta\rangle&=(-\varepsilon_{2}-\varepsilon_{1}\cos\phi)|\pi_{3}\rangle+\cos\phi|\eta_{n}\rangle-\sin\phi|\eta_{s}\rangle,\\
|\eta'\rangle&=-\varepsilon_{1}\sin\phi|\pi_{3}\rangle+\sin\phi|\eta_{n}\rangle+\cos\phi|\eta_{s}\rangle,
\end{split}
\label{zx}
\end{eqnarray}
One can define $\varepsilon=\varepsilon_{2}+\varepsilon_{1}cos\phi$, $\varepsilon'=\varepsilon_{1}sin\phi$. $\pi_{3}$  refer to the isospin $I=1$ component in the triplet.
We use the values of $\varepsilon=0.017 \pm 0.002$ , $\varepsilon'=0.004 \pm 0.001$, $\phi=39.0^\circ$ \cite{Kroll2005PP}.

For the $B_{s}$ meson function, we use the model \cite{AMLi2007,wangz2014}
\begin{eqnarray}
\phi_{B_s}(x,b)= N_{B_s}{x^2}(1-x)^2exp[-\frac {{M_{B_s}^2}x^2}{2{\omega_b}^2} - \frac 12 (\omega_b{b})^2],
\end{eqnarray}
where the normalization factor $N_{B_s}$ is dependent of the free parameter $\omega_b$.
$b$ is the conjugate variable of the parton transverse
momenta $k_{T}$. $M_{B_s}$ refers to the mass of the $B_s$ meson.
For the $B_{s}$ meson, one can obtain the value of $\omega_b = 0.50 \pm 0.05$
from the light cone sum rule \cite{hnl2003}.
In this paper, we will use those distribution amplitudes \cite{AMLi2007}:
\begin{eqnarray*}
\begin{split}
 \phi_{\pi}^A(x) &= \frac{3f_{\pi}}{\sqrt{6}} x(1-x)[ 1 +0.44C_2^{3/2}(t)],\\
 \phi_{\pi}^P(x) &=  \frac{f_{\pi}}{2\sqrt{6}}[1 +0.43C_2^{1/2}(t) ], \\
 \phi_{\pi}^T(x) &= -\frac{f_{\pi}}{2\sqrt{6}}[C_1^{1/2} (t)+0.55 C_3^{1/2} (t) ] ,\\
 \phi_{K}^A(x) &= \frac{3f_{K}}{\sqrt{6}}x(1-x)[1+0.17C_1^{3/2}(t)+0.2C_2^{3/2}(t)],\\
 \phi_{K}^P(x) &= \frac{f_{K}}{2\sqrt{6}} [1+0.24C_2^{1/2}(t)], \\
 \end{split}
\end{eqnarray*}
\begin{eqnarray}
\begin{split}
 \phi_{K}^T(x) &=-\frac{f_{K}}{2\sqrt{6}}[C_1^{1/2} (t)+0.35 C_3^{1/2} (t)] , \\
 \phi_{\phi} &=3\frac{f_{\phi}}{\sqrt{6}}x(1-x)[1+0.18 C_2^{3/2} (t)] ,
\end{split}
\end{eqnarray}
where $t=2x-1$. $f_{P(v)}$ are the decay constants of scalar (vector) mesons, respectively.
The pseudoscalar mesons $\pi,\eta$ and ${\eta'}$ have the similar wave functions.
The expressions of amplitudes can be obtained by the replacements
${\phi_\pi} \longrightarrow {\phi_\eta}, \;\;  {\phi_\pi}^p \longrightarrow {\phi_\eta}^p,  \;\;  {\phi_\pi}^t \longrightarrow {\phi_\eta}^t. $
Gegenbauer polynomials are defined as:
\begin{equation}
\begin{array}{ll}
 C^{1/2}_{1}(t)=t ,&C^{3/2}_{1}(t)=3t  \\
 C_2^{1/2}(t)=\frac{1}{2} (3t^2-1),& C_2^{3/2} (t)=\frac{3}{2}
(5t^2-1),
 \\
C_3^{1/2} (t) = \frac{1}{2} t (5t^2 -3)~.
\end{array}
\end{equation}

\subsection{\label{subsec:form}Calculation details}
In the framework of PQCD, we can calculate the $CP$ violation for the decay process  $\bar{B}_{s}\rightarrow P(V)\pi^{0}$ via $\pi^{0}-\eta-\eta'$ mixing. Firstly, we calculate the amplitudes $T$ and $P$, which can be decomposed in terms of tree and penguin contributions depending on the CKM matrix elements ${V_{ub}}{V_{ud}^*}$ and  ${V_{tb}}{V_{td}^*}$. Next, we take the decay process $\bar{B}_{s}\rightarrow K^{0}\pi^{0}$  and $\bar{B}_{s}\rightarrow \pi^{0}\eta(')$ as examples for the study of the $\pi^{0}-\eta-\eta'$ mixing mechanism.

\begin{center}
\textbf{1.   The $CP$ violation for the decay modes of $\bar{B}_{s}\rightarrow P(V)\pi^{0}$ except $\bar{B}_{s}\rightarrow \pi^{0}\eta(')$}
\end{center}

We take the decay process of $\bar{B}_{s}\rightarrow K^{0}\pi^{0}$ as example to introduce the $CP$ violation via $\pi^{0}-\eta-\eta'$ mixing. The decay amplitude A of $\bar{B}_{s}\rightarrow K^{0}\pi^{0}$  in PQCD can be written as
\begin{eqnarray}
\sqrt{2}A(\bar{B}_{s}\rightarrow K^{0}\pi^{0})=V_{ub}V^{*}_{ud}T_1-V_{tb}V^{*}_{td}P_1,
\end{eqnarray}
where $T_1$ and $P_1$ are the amplitudes form tree and penguin contributions, respectively. The tree level amplitude  $T_1$ can be given as
\begin{eqnarray}
T_1=f_{\pi} F_{B_s\to
K}^{LL} \left[a_{2}\right]+ M_{B_s\to K}^{LL} [C_{2}],
\label{t1}
\end{eqnarray}
and the penguin level amplitude $P_1$ can be written as
\begin{eqnarray}
P_1&=&f_{\pi} F_{B_s\to K}^{LL} \left[
 -a_{4}-\frac{3}{2}a_7+\frac{3}{2}a_9+\frac{1}{2}a_{10}\right]+f_{\pi} F_{B_s\to K}^{SP} \left[
 -a_{6}+\frac{1}{2}a_{8}\right] \nonumber\\
  &&
 +  M_{B_s\to K}^{LL} \left[-C_{3}+\frac{3}{2}c_8+\frac{1}{2}C_{9}+\frac{3}{2}c_{10}\right]
  +  f_{B_s}  F_{ann}^{SP}\left[-a_{6}+ \frac{1}{2}a_8\right]\nonumber\\
  &&
  +  f_{B_s} F_{ann}^{LL}\left[-a_{4}+\frac{1}{2}a_{10}\right] +  M_{ann}^{LL}\left[-C_{3}+\frac{1}{2}C_{9}\right]+  M_{ann}^{LR}\left[-C_{5}+\frac{1}{2}C_{7}\right],
\label{p1}
\end{eqnarray}
where the $f_{i}$ refers to the decay constant. The individual decay amplitudes in the above equations, such as $F_{B_s\to K}^{LL}$, $F_{B_s\to K}^{SP}$, $M_{B_s\to K}^{LL}$, $F_{ann}^{SP}$, $F_{ann}$, $M_{ann}^{LL}$ and $M_{ann}^{LR}$ arise from the $(V-A)(V-A)$, $(V-A)(V+A)$ and $(S-P)(S+P)$ operators, respectively, and will be given in Appendix.

Basing on the CKM matrix elements of $V_{ub}V^{*}_{ud}$ and $V_{tb}V^{*}_{td}$, we can express the decay amplitudes as following:
\begin{eqnarray}
\sqrt{2}A(\bar{B}_{s}\rightarrow\eta K^{0})&=&V_{ub}V^{*}_{ud}T_n-V_{tb}V^{*}_{td}P_n,\\
\sqrt{2}A(\bar{B}_{s}\rightarrow\eta' K^{0})&=&V_{ub}V^{*}_{ud}T_s-V_{tb}V^{*}_{td}P_s.
\end{eqnarray}
The contributions of $T_n$ and $P_n$ for the decay amplitudes $\bar{B}_{s}\rightarrow\eta K^{0}$ can be written as
\begin{eqnarray}
\begin{split}
   T_n &=f_n F_{B_s\to
K}^{LL} \left[a_{2}\right]+ M_{B_s\to K}^{LL} [C_{2}],
\end{split}
\end{eqnarray}
\begin{eqnarray}
\begin{split}
P_n &=f_n F_{B_s\to K}^{SP} \left[
 a_{6}-\frac{1}{2}a_8\right] +f_{n} F_{B_s\to K}^{LL} \left[
 2a_{3}+a_{4}-2a_{5}-\frac{1}{2}a_7+\frac{1}{2}a_9-\frac{1}{2}a_{10}\right]\\
  & \quad
  +f_{B_s} F_{ann}^{LL}\left[a_{4}-\frac{1}{2}a_{10}\right]
   +  M_{B_s\to K}^{LL} \left[C_{3}+2C_{4}+\frac{1}{2}C_8-\frac{1}{2}C_{9}+\frac{1}{2}C_{10}\right]   \\
   & \quad
  +  f_{B_s}  F_{ann}^{SP}\left[a_{6}- \frac{1}{2}a_8\right]
  +  M_{ann}^{LL}\left[C_{3}-\frac{1}{2}C_{9}\right]+  M_{ann}^{LR}\left[C_{5}-\frac{1}{2}C_{7}\right],
\end{split}
\end{eqnarray}
and
\begin{eqnarray}
\begin{split}
T_s &=0,
\end{split}
\end{eqnarray}
\begin{eqnarray}
\begin{split}
P_s &=f_{s} F_{B_s\to K}^{LL} \left[
 a_{3}-a_{5}+\frac{1}{2}a_7-\frac{1}{2}a_9\right] + f_{K} F_{B_s\to \eta_s}^{LL} \left[
 a_{4}-\frac{1}{2}a_{10}\right]  \\
 & \quad
 + f_{K} F_{B_s\to \eta_s}^{SP} \left[
 a_{6}-\frac{1}{2}a_{8}\right]+ M_{B_s\to K}^{LL} \left[C_{4}+C_{4}-\frac{1}{2}C_8-\frac{1}{2}C_{10}\right]  \\
  & \quad
  +  M_{B_s\to \eta_s}^{LL} \left[C_{3}-\frac{1}{2}C_{9}\right] +  M_{B_s\to \eta_s}^{LR} \left[C_{5}-\frac{1}{2}C_{7}\right] +  f_{B_s}  F_{ann}^{LL}\left[a_{4}- \frac{1}{2}a_{10}\right]  \\
  & \quad
  +  f_{B_s}  F_{ann}^{SP}\left[a_{6}- \frac{1}{2}a_{8}\right]
  +  M_{ann}^{LL}\left[C_{3}-\frac{1}{2}C_{9}\right]+  M_{ann}^{LR}\left[C_{5}-\frac{1}{2}C_{7}\right],
\end{split}
\end{eqnarray}
for the formula of $\sqrt{2}A(\bar{B}_{s}\rightarrow\eta' K^{0})=V_{ub}V^{*}_{ud}T_s-V_{tb}V^{*}_{td}P_s$.

The amplitudes T and P from the decay process of $\bar{B}_{s}\rightarrow K^{0}\pi^{0}$ with $\pi^{0}-\eta-\eta'$ mixing
can be written as:
\begin{eqnarray}
\begin{split}
 T &= T_1 + (\varepsilon_{1}+\varepsilon_{2}\cos\phi)T_n-\varepsilon_{2}\sin\phi T_s,\\
 P &= P_1 + (\varepsilon_{1}+\varepsilon_{2}\cos\phi)P_n-\varepsilon_{2}\sin\phi P_s.
 \label{tp}
\end{split}
\end{eqnarray}

One can see that the Eq.(\ref{tp}) without $\pi^{0}-\eta-\eta'$ mixing is reduced to
\begin{eqnarray}
 T = T_1,\;\;\;\;\;\;\; P = P_1,
 \end{eqnarray}
which are expressed in Eq.(\ref{t1}) and Eq.(\ref{p1}).

The relevant weak phase $\theta$ and strong phase $\delta$ are obtained as following
\begin{eqnarray}
r e^{i\delta} e^{i\theta}&=&\frac{P}{T}
\times\frac{V_{tb}V_{td}^*}{V_{ub}V_{ud}^*},  \label{eq:delform}
\end{eqnarray}
where the parameter $r$ represents the
absolute value of the ratio of penguin and tree amplitudes:
\begin{eqnarray}
r\equiv\Bigg|\frac{\langle K^{0}\pi^{0}|H^P|\bar{B}_{s}^{0}\rangle}{\langle K^{0}\pi^{0}|H^T|\bar{B}_{s}^{0}\rangle}\Bigg|
\label{r}.
\end{eqnarray}
The strong phase associated with $r$ can be given
\begin{eqnarray}
r e^{i\delta} &=&\frac{P}{T}
\times\bigg|\frac{V_{tb}V_{td}^*}{V_{ub}V_{ud}^*}\bigg| =r\cos\delta + \mathrm{i}r\sin\delta,  \label{eq:delform}
\end{eqnarray}
where
\begin{equation}
\left|\frac{V_{tb}V^{*}_{td}}{V_{ub}V^{*}_{ud}}\right|=\frac{\sqrt{[\rho(1-\rho)-\eta^2]^2+\eta^2}}{(1-\lambda^2/2)(\rho^2+\eta^2)}.
\label{3p}
\vspace{2mm}
\end{equation}
where $\rho$, $\eta$, $\lambda$ is the Wolfenstein parameters.

The $CP$ violation, $A_{CP}$, can be written as
\begin{eqnarray}
A_{CP}\equiv\frac{|A|^2-|\bar{A}|^2}{|A|^2+|\bar{A}|^2}=\frac{-2r
{\rm{sin}}\delta {\rm{sin}}\theta}{1+2r {\rm{cos}}\delta
{\rm{cos}}\theta+r^2}. \label{asy}
\end{eqnarray}

\begin{center}
\textbf{2.  The $CP$ violation of $\bar{B}_{s}\rightarrow \pi^{0}\eta(')$ via $\pi^{0}-\eta-\eta'$ mixing}
\end{center}

Due to the interference between $\pi^{0}$ and $\eta(')$, the effect of the isospin symmetry breaking is more significant for the decay process of $\bar{B}_{s}\rightarrow \pi^{0}\eta(')$. Hence, the $\pi^{0}-\eta-\eta'$ mixing, which including $\eta$ or $\eta'$ meson, may shift the phase larger so as to have a bigger impact on $CP$ violation.
 The decay amplitudes of $\bar{B}_{s}\rightarrow \pi^{0}\eta(')$ with isospin symmetry are defined as
\begin{eqnarray}
A(\bar{B}_{s}\rightarrow \pi^{0}\eta)&=&A(\bar{B}_{s}\rightarrow \pi^{0}\eta_{n})\cos\phi-A(\bar{B}_{s}\rightarrow \pi^{0}\eta_{s})\sin\phi,
\label{a}  \\ A(\bar{B}_{s}\rightarrow\pi^{0}\eta')&=&A(\bar{B}_{s}\rightarrow \pi^{0}\eta_{n})sin\phi+A(\bar{B}_{s}\rightarrow \pi^{0}\eta_{s})\cos\phi.  \label{eta}
\end{eqnarray}

Taking into account of $\pi^{0}-\eta-\eta'$ mixing, the decay amplitudes $\mathcal{A}$ for $\bar{B}_{s}\rightarrow \pi^{0}\eta$ in Eq.(\ref{a}) can be written as
\begin{eqnarray}
\mathcal{A}=\langle\pi^{0}\eta|{\mathcal{H}}_{eff}|\bar{B}_{s}\rangle = \langle\pi^{3}\eta|{\mathcal{H}}_{eff}|\bar{B}_{s}\rangle+(\varepsilon_{1}+\varepsilon_{2}\cos\phi)\langle\eta_{n}\eta|H|\bar{B}_{s}\rangle-\varepsilon_{2}\sin\phi\langle\eta_{s}
\eta|{\mathcal{H}}_{eff}|\bar{B}_{s}\rangle.
\label{ma}
\end{eqnarray}
We can define
\begin{eqnarray}
\begin{split}
{\mathcal{A}}_1&= \langle\pi^{3}\eta|{\mathcal{H}}_{eff}|\bar{B}_{s}\rangle \\
& =(-\varepsilon_{2}-\varepsilon_{1}\cos\phi)\langle\pi^{3}\pi^{3}|{\mathcal{H}}_{eff}|\bar{B}_{s}\rangle+\cos\phi\langle\pi^{3}\eta_{n}|{\mathcal{H}}_{eff}|\bar{B}_{s}\rangle-\sin\phi
\langle\pi^{3}\eta_{s}|{\mathcal{H}}_{eff}|\bar{B}_{s}\rangle,
\end{split}
\label{ma1}
\end{eqnarray}
\begin{eqnarray}
\begin{split}
{\mathcal{A}}_2& =(\varepsilon_{1}+\varepsilon_{2}\cos\phi)\langle\eta_{n}\eta|{\mathcal{H}}_{eff}|\bar{B}_{s}\rangle  \\
&  =(\varepsilon_{1}+\varepsilon_{2}\cos\phi)\cos\phi\langle\eta_{n}\eta_{n}|{\mathcal{H}}_{eff}|\bar{B}_{s}\rangle-(\varepsilon_{1}
+\varepsilon_{2}\cos\phi)\sin\phi\langle\eta_{n}\eta_{s}|{\mathcal{H}}_{eff}|\bar{B}_{s}\rangle
+\mathcal{O} (\varepsilon),
\end{split}
\label{ma2}
\end{eqnarray}
and
\begin{eqnarray}
\begin{split}
{\mathcal{A}}_3& =-\varepsilon_{2}\sin\phi\langle\eta_{s}\eta|{\mathcal{H}}_{eff}|\bar{B}_{s}\rangle  \\
&
=-\varepsilon_{2}\sin\phi\cos\phi\langle\eta_{n}\eta_{s}|{\mathcal{H}}_{eff}|\bar{B}_{s}\rangle+\varepsilon_{2}\sin\phi\cos\phi\langle\eta_{s}\eta_{s}
|{\mathcal{H}}_{eff}|\bar{B}_{s}\rangle+\mathcal{O} (\varepsilon),
\end{split}
\label{ma3}
\end{eqnarray}
where
\begin{eqnarray}
\mathcal{O} (\varepsilon)=\mathcal{O} (\varepsilon_1)+\mathcal{O} (\varepsilon_2),
\end{eqnarray}
and we have ignored the higher order term of $\varepsilon$. One can express $\mathcal{A}= {\mathcal{A}}_1 + {\mathcal{A}}_2 + {\mathcal{A}}_3$.

In the same way, we can present the decay amplitudes $\mathcal{A}'$ for $\bar{B}_{s}\rightarrow \pi^{0}\eta'$ with $\pi^{0}-\eta-\eta'$ mixing in Eq.(\ref{eta})
\begin{eqnarray}
\begin{split}
{\mathcal{A}}'&
=\langle\pi^{0}\eta'|{\mathcal{H}}_{eff}|\bar{B}_{s}\rangle
\\
&
=\langle\pi^{3}\eta'|{\mathcal{H}}_{eff}|\bar{B}_{s}\rangle+(\varepsilon_{1}+\varepsilon_{2}\cos\phi)\langle\eta_{n}\eta'|{\mathcal{H}}_{eff}|\bar{B}_{s}
\rangle-\varepsilon_{2}\sin\phi\langle\eta_{s}\eta'|{\mathcal{H}}_{eff}|\bar{B}_{s}\rangle
\\
&
=-\varepsilon_{1}\sin\phi\langle\pi^{3}\pi^{3}|{\mathcal{H}}_{eff}|\bar{B}_{s}\rangle+ \sin\phi\langle\pi^{3}\eta_{n}|{\mathcal{H}}_{eff}|\bar{B}_{s}\rangle+\cos\phi\langle\pi^{3}\eta_{s}|{\mathcal{H}}_{eff}|\bar{B}_{s}\rangle \\
& \quad
+(\varepsilon_{1}+\varepsilon_{2}\cos\phi)\sin\phi\langle\eta_{n}\eta_{n}|{\mathcal{H}}_{eff}|\bar{B}_{s}\rangle+(\varepsilon_{1}+\varepsilon_{2}\cos\phi)
\cos\phi\langle\eta_{n}\eta_{s}|{\mathcal{H}}_{eff}|\bar{B}_{s}\rangle
\\
& \quad
-\varepsilon_{2}\sin\phi\sin\phi\langle\eta_{s}\eta_{n}|{\mathcal{H}}_{eff}|\bar{B}_{s}\rangle-\varepsilon_{2}\sin\phi\cos\phi\langle\eta_{s}\eta_{s}|{\mathcal{H}}_{eff}|\
\bar{B}_{s}\rangle.
\label{maa}
\end{split}
\end{eqnarray}

Hence, depending on the CKM matrix elements $V_{ub}V^{*}_{ud}$ and $V_{tb}V^{*}_{td}$ , we can express the decay amplitudes $A(\bar{B}_{s}\rightarrow \pi^{0}\eta(')$ as following:
\begin{eqnarray}
A(\bar{B}_{s}\rightarrow \pi^{0}\eta^{({'})})=V_{ub}V^{*}_{ud}T - V_{tb}V^{*}_{td}P,
\label{aa}
\end{eqnarray}
Where $T$ and $P$ refer to the tree and penguin contributions from $\mathcal{A}$ and ${\mathcal{A}}'$ in Eq.(\ref{ma}),(\ref{maa}), respectively. The relevant amplitudes can be obtained from the decay processes of $\bar{B}_{s}\rightarrow \pi^{0}\pi^{0}$, $\bar{B}_{s}\rightarrow \pi^{0}\eta_{n}$, $\bar{B}_{s}\rightarrow \pi^{0}\eta_{s}$, $\bar{B}_{s}\rightarrow \eta_{n}\eta_{n}$ and $\bar{B}_{s}\rightarrow \eta_{s}\eta_{s}$.
Combined with Eq.(\ref{ma}), (\ref{ma1}), (\ref{ma2}), (\ref{ma3}), (\ref{maa}), (\ref{aa}) we can also obtain $CP$ violation from the Eqs.(\ref{r}), (\ref{eq:delform}), (\ref{3p}) and (\ref{asy}).

\section{\label{input}INPUT PARAMETERS}
The CKM matrix, which elements are determined from experiments, can be expressed in terms of the Wolfenstein parameters $A$, $\rho$, $\lambda$ and $\eta$ \cite{wol}:
\begin{equation}
\left(
\begin{array}{ccc}
  1-\tfrac{1}{2}\lambda^2   & \lambda                  &A\lambda^3(\rho-\mathrm{i}\eta) \\
  -\lambda                 & 1-\tfrac{1}{2}\lambda^2   &A\lambda^2 \\
  A\lambda^3(1-\rho-\mathrm{i}\eta) & -A\lambda^2              &1\\
\end{array}
\right),\label{ckm}
\end{equation}
where $\mathcal{O} (\lambda^{4})$ corrections are neglected. The latest values for the parameters in the CKM matrix are \cite{PDG2018}:
\begin{eqnarray}
&& \lambda=0.22506\pm0.00050,\quad A=0.811\pm0.026,\nonumber \\
&& \bar{\rho}=0.124_{-0.018}^{+0.019},\quad
\bar{\eta}=0.356\pm{0.011}.\label{eq: rhobarvalue}
\end{eqnarray}
where
\begin{eqnarray}
 \bar{\rho}=\rho(1-\frac{\lambda^2}{2}),\quad
\bar{\eta}=\eta(1-\frac{\lambda^2}{2}).\label{eq: rho rhobar
relation}
\end{eqnarray}
From Eqs. (\ref{eq: rhobarvalue}) ( \ref{eq: rho rhobar relation})
we have
\begin{eqnarray}
0.109<\rho<0.147,\quad  0.354<\eta<0.377.\label{eq: rho value}
\end{eqnarray}
The other parameters are given as following \cite{wol,PDG2018}:
\begin{eqnarray}
f_{\pi}&=&0.13\text{GeV}, \hspace{2.95cm}  f_{K}=0.16\text{GeV},        \nonumber \\
m_{B^0_s}&=&5.37\text{GeV},\hspace{2.95cm} f_{B_s}=0.23\text{GeV},   \nonumber \\
f_n&=&0.17\text{GeV},\hspace{2.95cm} f_s=0.14\text{GeV},  \nonumber \\
m_\pi&=&0.14\text{GeV},\hspace{2.95cm} m_W=80.39\text{GeV},  \nonumber \\
m_t&=&173.21\text{GeV},\hspace{2.95cm} m_b=4.8\text{MeV}.
\end{eqnarray}

\section{\label{num}Numerical results}

 The CP violation depends on the weak phase differences from the CKM matrix elements and the strong phase differences associated with QCD.
The CKM matrix elements are determined by the parameters of  $A$, $\rho$, $\lambda$ and $\eta$.
We find that the results for the $CP$ violation are less reliant on $A$ and $\lambda$ in the course of calculations.
Hence, we present the $CP$ violation from the weak phases associated with the $\rho$ and $\eta$ in the CKM matrix elements
while the $A$ and $\lambda$ are assigned for the central values. In Table.I, we show the values of $CP$ violation of $B_s$ decay modes from isospin symmetry and isospin symmetry breaking via $\pi^{0}-\eta-\eta^{'}$ mixing. From Table.I, it can be seen that the increasing rate of the $CP$ violation, which are defined ${\frac{|{x_2}| - |{x_1}|}{|x_1|}}\times 100\%$ (where $x_1$, $x_2$ represent the $CP$ violation values from isospin symmetry and isospin symmetry breaking, respectively.), is larger
in $\bar{B}_{s}\rightarrow \pi^{0}\eta(')$ decay process comparing with the other decay channels we are considering.
It is intelligible that the final states for the decay process include the $\eta$ or $\eta'$ meson. Due to the isospin symmetry breaking, the interference between the $\pi^{0}$ and $\eta(')$ mesons is stronger than other decay channels whose final states don't contain $\eta$ or $\eta'$ meson. Hence, these decay channels including $\eta$ or $\eta'$ meson make the strong phase larger resulting in a great impact on $CP$ violation. We can find that the $CP$ violation of the decay mode $\bar{B}_{s}\rightarrow K^{0}\pi^{0}$ has not changed much in Table.I. The $CP$ violation of the decay mode $\bar{B}_{s}\rightarrow K^{0*}\pi^{0}$ is changed from $ -23.58 $$\%$ to $ -36.88 $$\%$. From Table.I, one can also see that the isospin symmetry breaking changes the sign of the $CP$ violation, for example from $ 8.32 $$\%$ to $ -6.33 $$\%$ for the decay channel
of $\bar{B}_{s}\rightarrow \phi\pi^{0}$, from $ -8.76 $$\%$ to $ 15.48 $$\%$ for the decay channel of $\bar{B}_{s}\rightarrow \pi^{0}\eta$. The increasing rate of $CP$ violation for the decay mode $\bar{B}_{s}\rightarrow \pi^{0}\eta^{'}$ is $-65.99$$\%$ for the central value.
\tabcolsep 0.33in
\begin{table}
\centering
\caption{The CP violation of $B_s$ decay mode via isospin symmetry and isospin symmetry breaking via $\pi^{0}-\eta-\eta^{'}$ mixing.
The increasing rate is defined ${\frac{|{x_2}| - |{x_1}|}{|x_1|}}\times 100\%$,
where $x_1$, $x_2$ represent the values of  $CP$ violation($\%$) from isospin symmetry and isospin symmetry breaking, respectively.
The fluctuation numerical values refer to the contribution of the limiting parameters from the CKM matrix elements.}
\begin{tabular}[t]{lccc}
\hline\hline \\
\multicolumn{1}{c}{decay mode} & isospin symmetry &$\pi^{0}-\eta-\eta^{'}$ mixing& the increasing rate\\ \\
\cline{1-4} \\
$\bar{B}_{s}\rightarrow K^{0}\pi^{0}$
&$ 53.43_{-2.14}^{+2.26} $ $\%$& $ 51.44_{-3.00}^{+3.30} $ $\%$& $ -3.72_{-1.87}^{+2.01}$  $\%$\\ \\
$\bar{B}_{s}\rightarrow K^{0*}\pi^{0}$
&$ -23.58_{-1.22}^{+1.12} $ $\%$& $ -36.88_{-0.70}^{+0.73} $ $\%$& $56.4_{-4.87}^{+4.55}$  $\%$\\ \\
 $\bar{B}_{s}\rightarrow \phi\pi^{0}$
&$ 8.32_{-0.47}^{+0.48} $ $\%$& $ -6.33_{-0.20}^{+0.18} $ $\%$& $-23.92_{-6.42}^{+6.85}$  $\%$\\ \\
 $\bar{B}_{s}\rightarrow \pi^{0}\eta$
&$ -8.76_{-0.25}^{+0.28} $ $\%$& $ 15.48_{-0.73}^{+0.81} $ $\%$& $ 76.71_{-13.00}^{+15.39}$  $\%$\\ \\
$\bar{B}_{s}\rightarrow \pi^{0}\eta^{'}$
&$ 27.43 _{-1.05}^{+1.09}$ $\%$& $ 9.33_{-0.60}^{+0.67} $ $\%$& $-65.99_{-0.92}^{+1.05}$  $\%$\\ \\
\hline\hline
\end{tabular}
\end{table}

From Table.I, we can find great changes between the values of $CP$ violation from isospin symmetry and isospin symmetry breaking via $\pi^{0}-\eta-\eta^{'}$ mixing. In order to study the influence of the weak phase on $CP$ violation and understand the $\pi^{0}-\eta-\eta^{'}$ mixing mechanism, we present the $CP$ violation as a function of $\rho$ and $\eta$ in Fig.1 while taking the mixing parameters $\varepsilon_{1}$ and $\varepsilon_{2}$ as central values. We vary $(\rho,\eta)$
from the limiting values $(\rho_{min},\eta_{min})$ to $(\rho_{max},\eta_{max})$, respectively, in Fig.1. Due to the effect of weak phases from CKM matrix elements, the value of $CP$ violation for the decay process of $\bar{B}_{s}\rightarrow \pi^{0}\eta$ changes from $ 14.75 $$\%$ to $ 16.29 $$\%$ taking into account of isospin symmetry breaking.

\begin{figure}
\centering
\includegraphics[height=5cm,width=6cm]{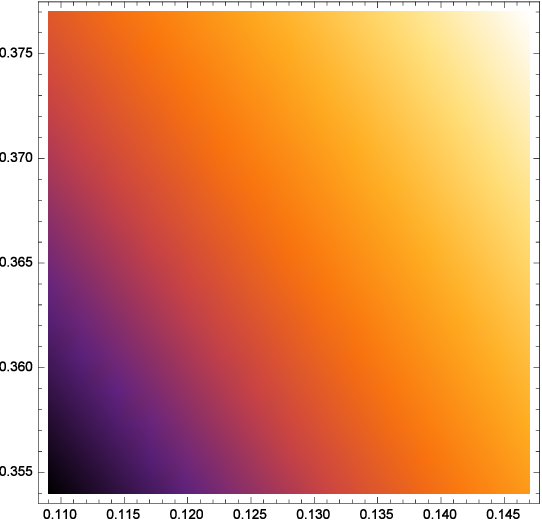}
\includegraphics[height=5cm,width=1cm]{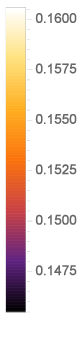}
\caption{The direct $CP$ violation as a function of $\rho$ and $\eta$ from the CKM matrix element with isospin symmetry breaking
for the decay process of $\bar{B}_{s}\rightarrow \pi^{0}\eta$.
The horizontal axis and vertical axis refer to the values of $\rho$ and $\eta$, respectively.}
\label{fig1}
\end{figure}
It can be seen from the Eq.(\ref{asy}) that the value of direct $CP$ violation is also dependent on $\sin\delta$ and $r$. We take the decay channel of $\bar{B}_{s}\rightarrow \pi^{0}\eta$ as an example. When $\rho$ and $\eta$ are taken as the central value for the CKM matrix elements, we present the direct $CP$ violation as a function of $\varepsilon_{1}$, $\varepsilon_{2}$ in Fig.2a from the isospin symmetry and in Fig.2b from isospin symmetry breaking via $\pi^{0}-\eta-\eta^{'}$ mixing. Comparing the Fig.2a with the Fig.2b, the $CP$ violation value has a great change.
Only considering the central value, the value of $CP$ violation changes from $ -8.76 $$\%$ in Fig.2a to $ 15.48 $$\%$ in Fig.2b and shifts the sign. In Fig.3 and Fig.4, we give the numerical result of $\sin\delta$ and $r$ for the decay process of $\bar{B}_{s}\rightarrow \pi^{0}\eta$. Comparing Fig.3a with Fig.3b, we can find that the value of $\sin\delta$ changes the sign from $0.276$ in Fig.3a to the central value $-2.219$ in Fig.3b. In Fig.4, the central value of $r$ changes large comparing the result of isospin symmetry breaking in Fig.4b to the value from isospin symmetry in Fig.4a. Based on the changes of $\sin\delta$ and $r$, large $CP$ violation is obtained from isospin symmetry breaking via $\pi^{0}-\eta-\eta^{'}$ mixing.

\begin{figure*}[t]
\centering
\subfigure[ ]{
  \includegraphics[height=5cm,width=6cm]{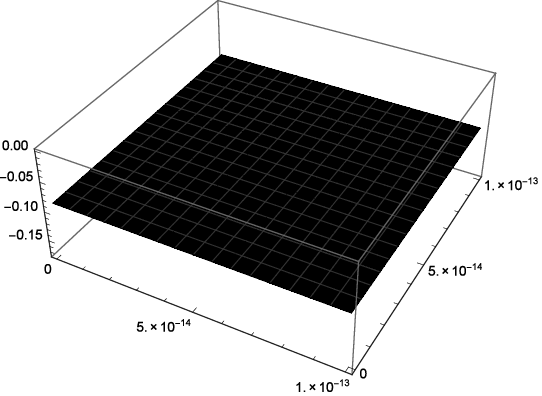}
 \includegraphics[height=5cm,width=1cm]{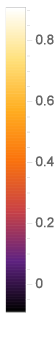} }
\hfill
\centering
\subfigure[]{
  \includegraphics[height=5cm,width=6cm]{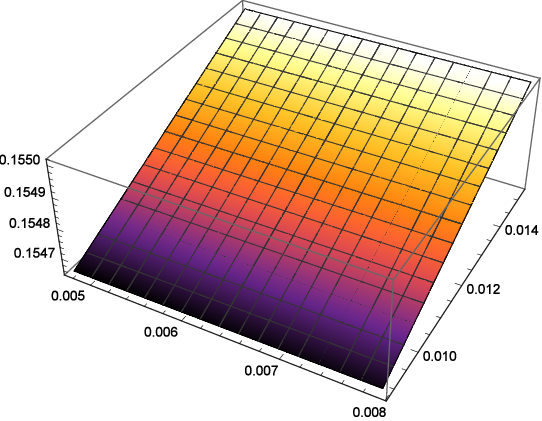}
 \includegraphics[height=5cm,width=1cm]{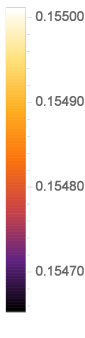}}
\caption{(a)The direct $CP$ violation as a function of $\varepsilon_{1}$ and $\varepsilon_{2}$ from the effects of isospin symmetry
for the decay process of $\bar{B}_{s}\rightarrow \pi^{0}\eta$.
(b)The same as (a) from the effects of isospin symmetry breaking.
The horizontal axis and vertical axis refer to the values of $\varepsilon_{1}$ and $\varepsilon_{2}$, respectively.}
\end{figure*}

\begin{figure*}[t]
\centering
\subfigure[ ]{
  \includegraphics[height=5cm,width=6cm]{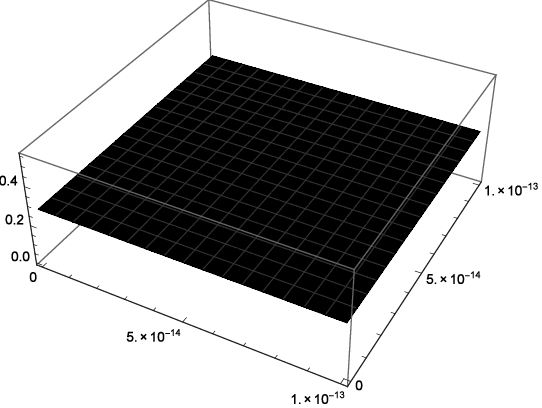}
 \includegraphics[height=5cm,width=1cm]{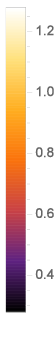} }
\hfill
\centering
\subfigure[ ]{
  \includegraphics[height=5cm,width=6cm]{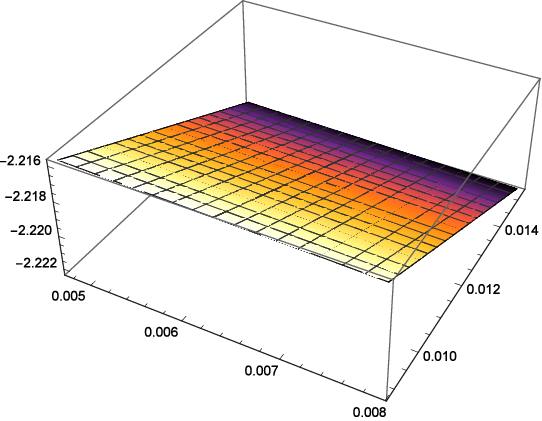}
 \includegraphics[height=5cm,width=1cm]{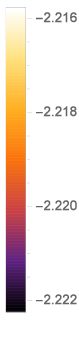}}
\caption{(a)The value of $\sin\delta$  as a function of $\varepsilon_{1}$ and $\varepsilon_{2}$ from the effects of the isospin symmetry
for the decay process of $\bar{B}_{s}\rightarrow \pi^{0}\eta$.
(b)The same as (a) from the effects of isospin symmetry breaking.
The horizontal axis and vertical axis refer to the values of $\varepsilon_{1}$ and $\varepsilon_{2}$, respectively.}
\end{figure*}

\begin{figure*}[t]
\centering
\subfigure[ ]{
  \includegraphics[height=5cm,width=6cm]{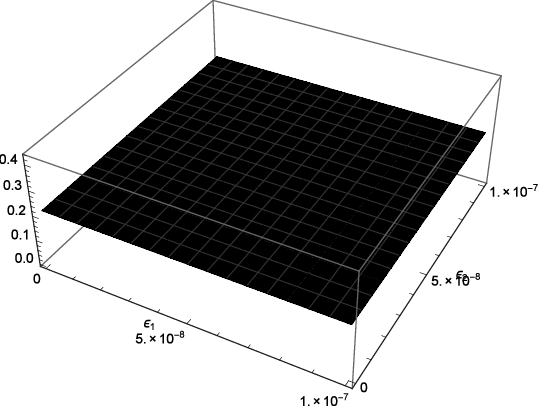}
 \includegraphics[height=5cm,width=1cm]{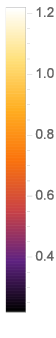} }
\hfill
\centering
\subfigure[ ]{
  \includegraphics[height=5cm,width=6cm]{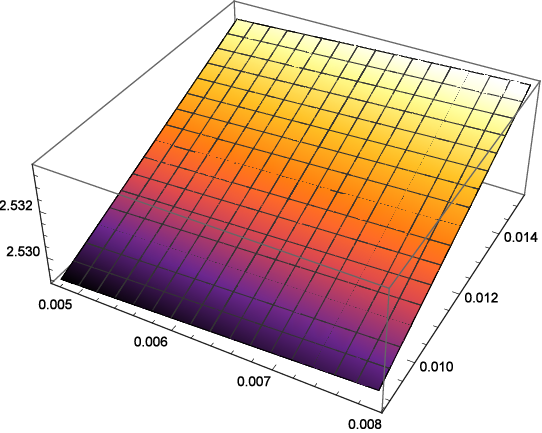}
 \includegraphics[height=5cm,width=1cm]{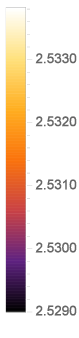}}
\caption{(a)The value of $r$  as a function of $\varepsilon_{1}$ and $\varepsilon_{2}$ from the effects of isospin symmetry
for the decay process of $\bar{B}_{s}\rightarrow \pi^{0}\eta$.
(b)The same as (a) from the effects of isospin symmetry breaking.
The horizontal axis and vertical axis refer to the values of $\varepsilon_{1}$ and $\varepsilon_{2}$, respectively.}
\end{figure*}

\section{\label{sum}SUMMARY AND CONCLUSION}
In this paper, we study the $CP$ violation for the decay process of $\bar{B}_{s}\rightarrow P(V)\pi^{0}$ in Perturbative QCD.
It is found that the $CP$ violation can be shifted  via $\pi^{0}-\eta-\eta'$ mixing from the isospin symmetry breaking.
The $CP$ violation arises from the weak
phase difference in CKM matrix and the strong phase difference.
The $CP$ violation changes small for the decay mode $\bar{B}_{s}\rightarrow K^{0}\pi^{0}$ via $\pi^{0}-\eta-\eta'$ mixing and the central value of the increasing rate is $-3.72$$\%$.
The rate of increase of the $CP$ violation is larger
for the decay process of $\bar{B}_{s}\rightarrow \pi^{0}\eta(')$ than other decay channels. This is due to the breaking of isospin symmetry, the interference between the $\pi^{0}$ and $\eta(')$ mesons is strong than other decay channels which final state doesn't contain $\eta$ or $\eta'$ meson.
For the decay process $\bar{B}_{s}\rightarrow \phi\pi^{0}$ and $\bar{B}_{s}\rightarrow \pi^{0}\eta$,
the isospin symmetry breaking changes the sign of the $CP$ violation.

In order to achieve the required energy and luminosity requirements, the Large Hadron Collider (LHC), which has currently started at CERN, has been upgraded many times.  The LHC Run \uppercase\expandafter{\romannumeral1} data started in 2010. The peak instantaneous luminosity documentary during Run \uppercase\expandafter{\romannumeral1} was $8.0\times10^{32} \text{cm}^{-2}\text{s}^{-1}$. The center-of mass energy was primarily $\sqrt{s}$ = 7 TeV and was raised to 8 TeV in 2012 \cite{Aaij:2014jba}. This was followed by the first long shutdown period (LS1), which was devoted to upgrades essential for increasing beam energy to $\sqrt{s}$= 13 TeV centre of mass energy and peak instantaneous luminosity $1.7\times10^{34} \text{cm}^{-2}\text{s}^{-1}$ \cite{Szumlak:2017und,Whitehead:2017vob}. In the following years, there are two primary detector (CMS and ATLAS) upgrades happening after Run \uppercase\expandafter{\romannumeral2} and Run \uppercase\expandafter{\romannumeral3}. Phase-\uppercase\expandafter{\romannumeral1} and \uppercase\expandafter{\romannumeral2} upgrade prepares for an instantaneous luminosity of $2-3\times10^{34} \text{cm}^{-2}\text{s}^{-1}$ and $5-7\times10^{34} \text{cm}^{-2}\text{s}^{-1}$ \cite{Chen:2017khz}, respectively.
 With a series of modifications and upgrades, the LHC gives access to high energy frontier at TeV scale and an occasion to further improve the consistency test for the CKM matrix. The production rates for heavy quark flavors will be great at the LHC, and the $b \bar b$ production cross section will be of the order of 0.5 mb, providing as many as $10^{12}$ bottom events per year \cite{Aaij:2014jba,Schopper:2006he}. The heavy quark physics is one of the major topics of LHC experiments. Especially, the LHCb experiment exploits amounts of $b$ mesons, produced in proton-proton collisions at the LHC to search for $CP$ violation.
Recently, LHCb Collaboration presents observation of the decay $B_{s}^{0}\rightarrow \phi\pi^{+}\pi^{-}$ meson.
Obtaining more data from LHC, it is possible to make further analysis for $CP$ violation of $B_{s}^{0}$ decays \cite{LHC2017}.
We expect that our results is valuable for measurement of $CP$ violation of $B_{s}^{0}$ decays in the following LHCb experiments.

\section{Acknowledgments}
This work was supported by National Natural Science
Foundation of China (Project Numbers 11605041),
and the Research Foundation of the young core teacher from Henan province.

\section{APPENDIX: Related functions defined in the text}
The functions related with the tree and penguin contributions are presented for the factorization and non-factorization
amplitudes with PQCD approach \cite{YA,LKM,AMLi2007}.

The hard scales $t$ are chosen as \begin{eqnarray}
t_a&=&\mbox{max}\{{\sqrt
{x_3}M_{B_s},1/b_1,1/b_3}\},\\
t_a^\prime&=&\mbox{max}\{{\sqrt
{x_1}M_{B_s},1/b_1,1/b_3}\},\\
t_b&=&\mbox{max}\{\sqrt
{x_1x_3}M_{B_s},\sqrt{|1-x_1-x_2|x_3}M_{B_s},1/b_1,1/b_2\},\\
t_b^\prime&=&\mbox{max}\{\sqrt{x_1x_3}M_{B_s},\sqrt
{|x_1-x_2|x_3}M_{B_s},1/b_1,1/b_2\},\\
t_c&=&\mbox{max}\{\sqrt{1-x_3}M_{B_s},1/b_2,1/b_3\},\\
t_c^\prime
&=&\mbox{max}\{\sqrt {x_2}M_{B_s},1/b_2,1/b_3\},\\
t_d&=&\mbox{max}\{\sqrt {x_2(1-x_3)}M_{B_s},
\sqrt{1-(1-x_1-x_2)x_3}M_{B_s},1/b_1,1/b_2\},\\
t_d^\prime&=&\mbox{max}\{\sqrt{x_2(1-x_3)}M_{B_s},\sqrt{|x_1-x_2|(1-x_3)}M_{B_s},1/b_1,1/b_2\}.
\end{eqnarray}

The function $h$ comprises the jet function $S_t(x_i)$ arising from the threshold
re-summation\cite{L3} and the propagator of virtual quark
and gluon \cite{YA,LKM,AMLi2007}.
They are defined by
\begin{eqnarray}
h_e(x_1,x_3,b_1,b_3)&=&\left[\theta(b_1-b_3)I_0(\sqrt
x_3M_{B_s}b_3)K_0(\sqrt
x_3 M_{B_s}b_1)\right.\\
&& \left.+\theta(b_3-b_1)I_0(\sqrt x_3M_{B_s}b_1)K_0(\sqrt
x_3M_{B_s}b_3)\right]K_0(\sqrt {x_1x_3}M_{B_s}b_1)S_t(x_3),\nonumber\\
h_n(x_1,x_2,x_3,b_1,b_2)&=&\left[\theta(b_2-b_1)K_0(\sqrt
{x_1x_3}M_{B_s}b_2)I_0(\sqrt
{x_1x_3}M_{B_s}b_1)\right. \nonumber\\
&&\;\;\;\left. +\theta(b_1-b_2)K_0(\sqrt
{x_1x_3}M_{B_s}b_1)I_0(\sqrt{x_1x_3}M_{B_s}b_2)\right]\nonumber\\
&&\times
\left\{\begin{array}{ll}\frac{i\pi}{2}H_0^{(1)}(\sqrt{(x_2-x_1)x_3}
M_{B_s}b_2),& x_1-x_2<0\\
K_0(\sqrt{(x_1-x_2)x_3}M_{B_s}b_2),& x_1-x_2>0
\end{array}
\right. ,
\end{eqnarray}
\begin{eqnarray}
h_a(x_2,x_3,b_2,b_3)&=&(\frac{i\pi}{2})^2
S_t(x_3)\Big[\theta(b_2-b_3)H_0^{(1)}(\sqrt{x_3}M_{B_s}b_2)J_0(\sqrt
{x_3}M_{B_s}b_3)\nonumber\\
&&\;\;+\theta(b_3-b_2)H_0^{(1)}(\sqrt {x_3}M_{B_s}b_3)J_0(\sqrt
{x_3}M_{B_s}b_2)\Big]H_0^{(1)}(\sqrt{x_2x_3}M_{B_s}b_2),
\\
h_{na}(x_1,x_2,x_3,b_1,b_2)&=&\frac{i\pi}{2}\left[\theta(b_1-b_2)H^{(1)}_0(\sqrt
{x_2(1-x_3)}M_{B_s}b_1)J_0(\sqrt {x_2(1-x_3)}M_{B_s}b_2)\right. \nonumber\\
&&\;\;\left.
+\theta(b_2-b_1)H^{(1)}_0(\sqrt{x_2(1-x_3)}M_{B_s}b_2)J_0(\sqrt
{x_2(1-x_3)}M_{B_s}b_1)\right]\nonumber\\
&&\;\;\;\times K_0(\sqrt{1-(1-x_1-x_2)x_3}M_{B_s}b_1),
\\
h_{na}^\prime(x_1,x_2,x_3,b_1,b_2)&=&\frac{i\pi}{2}\left[\theta(b_1-b_2)H^{(1)}_0(\sqrt
{x_2(1-x_3)}M_{B_s}b_1)J_0(\sqrt{x_2(1-x_3)}M_{B_s}b_2)\right. \nonumber\\
&&\;\;\;\left. +\theta(b_2-b_1)H^{(1)}_0(\sqrt
{x_2(1-x_3)}M_{B_s}b_2)J_0(\sqrt{x_2(1-x_3)}M_{B_s}b_1)\right]\nonumber\\
&&\;\;\;\times
\left\{\begin{array}{ll}\frac{i\pi}{2}H^{(1)}_0(\sqrt{(x_2-x_1)(1-x_3)}M_{B_s}b_1),&
x_1-x_2<0\\
K_0(\sqrt {(x_1-x_2)(1-x_3)}M_{B_s}b_1),&
x_1-x_2>0\end{array}\right. ,
\end{eqnarray}
where $H_0^{(1)}(z) = \mathrm{J}_0(z) + i\, \mathrm{Y}_0(z)$.

The $S_t$ re-sums the threshold logarithms $\ln^2x$ appearing in the
hard kernels to all orders and it has been parameterized as
  \begin{eqnarray}
S_t(x)=\frac{2^{1+2c}\Gamma(3/2+c)}{\sqrt \pi
\Gamma(1+c)}[x(1-x)]^c,
\end{eqnarray}
with $c=0.4$. In the nonfactorizable contributions, $S_t(x)$ gives
a very small numerical effect on the amplitude~\cite{L4}.
Therefore, we drop $S_t(x)$ in $h_n$ and $h_{na}$.

The evolution factors $E^{(\prime)}_e$ and $E^{(\prime)}_a$ are given by \cite{YA,LKM,AMLi2007}
\begin{eqnarray}
E_e(t)&=&\alpha_s(t) \exp[-S_B(t)-S_3(t)],
 \ \ \ \
 E'_e(t)=\alpha_s(t)
 \exp[-S_B(t)-S_2(t)-S_3(t)]|_{b_1=b_3},\\
E_a(t)&=&\alpha_s(t)
 \exp[-S_2(t)-S_3(t)],\
 \ \ \
E'_a(t)=\alpha_s(t) \exp[-S_B(t)-S_2(t)-S_3(t)]|_{b_2=b_3},
\end{eqnarray}
in which the Sudakov exponents are defined as
\begin{eqnarray}
S_B(t)&=&s\left(x_1\frac{M_{B_s}}{\sqrt
2},b_1\right)+\frac{5}{3}\int^t_{1/b_1}\frac{d\bar \mu}{\bar
\mu}\gamma_q(\alpha_s(\bar \mu)),\\
S_2(t)&=&s\left(x_2\frac{M_{B_s}}{\sqrt
2},b_2\right)+s\left((1-x_2)\frac{M_{B_s}}{\sqrt
2},b_2\right)+2\int^t_{1/b_2}\frac{d\bar \mu}{\bar
\mu}\gamma_q(\alpha_s(\bar \mu)),
\end{eqnarray}
 with the quark
anomalous dimension $\gamma_q=-\alpha_s/\pi$. Replacing the
kinematic variables of $M_2$ to $M_3$ in $S_2$, we can get the
expression for $S_3$. The explicit form for the  function
$s(Q,b)$ is \cite{YA,LKM,AMLi2007}:
\begin{eqnarray}
s(Q,b)&=&~~\frac{A^{(1)}}{2\beta_{1}}\hat{q}\ln\left(\frac{\hat{q}}
{\hat{b}}\right)-
\frac{A^{(1)}}{2\beta_{1}}\left(\hat{q}-\hat{b}\right)+
\frac{A^{(2)}}{4\beta_{1}^{2}}\left(\frac{\hat{q}}{\hat{b}}-1\right)
-\left[\frac{A^{(2)}}{4\beta_{1}^{2}}-\frac{A^{(1)}}{4\beta_{1}}
\ln\left(\frac{e^{2\gamma_E-1}}{2}\right)\right]
\ln\left(\frac{\hat{q}}{\hat{b}}\right)
\nonumber \\
&&+\frac{A^{(1)}\beta_{2}}{4\beta_{1}^{3}}\hat{q}\left[
\frac{\ln(2\hat{q})+1}{\hat{q}}-\frac{\ln(2\hat{b})+1}{\hat{b}}\right]
+\frac{A^{(1)}\beta_{2}}{8\beta_{1}^{3}}\left[
\ln^{2}(2\hat{q})-\ln^{2}(2\hat{b})\right],
\end{eqnarray} where the variables are defined by
\begin{eqnarray}
\hat q\equiv \mbox{ln}[Q/(\sqrt 2\Lambda)],~~~ \hat b\equiv
\mbox{ln}[1/(b\Lambda)], \end{eqnarray} and the coefficients
$A^{(i)}$ and $\beta_i$ are \begin{eqnarray}
\beta_1=\frac{33-2n_f}{12},~~\beta_2=\frac{153-19n_f}{24},\nonumber\\
A^{(1)}=\frac{4}{3},~~A^{(2)}=\frac{67}{9}
-\frac{\pi^2}{3}-\frac{10}{27}n_f+\frac{8}{3}\beta_1\mbox{ln}(\frac{1}{2}e^{\gamma_E}),
\end{eqnarray}
$n_f$ is the number of the quark flavors and $\gamma_E$ is the
Euler constant. We will use the one-loop running coupling
constant, i.e. we pick up the four terms in the first line of the
expression for the function $s(Q,b)$ \cite{YA,LKM,AMLi2007}.

The $LL$, $LR$ and $SP$ refer to the contributions from
$(V-A)(V-A)$ operators, $(V-A)(V+A)$ operators and $(S-P)(S+P)$ operators, respectively.
The form factor of $B_{s}\rightarrow M_3$ can be given \cite{YA,LKM,AMLi2007}:
\begin{itemize}
\item $(V-A)(V-A)$ operators:
  \begin{eqnarray}
  f_{M_2} F^{LL}_{B_s\to M_3} (a_i)&=&8\pi
  C_FM_{B_s}^4f_{M_2}\int^1_0dx_1dx_3\int^\infty_0b_1db_1b_3db_3
\phi_{B_s}(x_1,b_1)
  \Big\{a_i(t_a) E_e(t_a)
  \nonumber\\
  &&\times \Big[(1+x_3)\phi_3^A(x_3)+r_3(1-2x_3)(\phi_3^P(x_3)+\phi_3^T(x_3))
  \Big]h_e(x_1,x_3,b_1,b_3)
 \nonumber\\ && \;\;+2r_3\phi_3^P(x_3)a_i(t_a^\prime) E_e(t_a^\prime)h_e(x_3,x_1,b_3,b_1)
 \Big\},\label{ppefll}
  \end{eqnarray}

\item $(V-A)(V+A)$ operators:
\begin{eqnarray}
  F^{LR}_{B_s\to M_3}(a_i)&=&-F^{LL}_{B_s\to M_3}(a_i),\label{ppeflr}
\end{eqnarray}

\item $(S-P)(S+P)$ operators:
  \begin{eqnarray}
 f_{M_2} F^{SP}_{B_s\to M_3}(a_i)&=& 16\pi r_2
  C_FM_{B_s}^4f_{M_2}\int^1_0dx_1dx_3\int^\infty_0b_1db_1b_3db_3
\phi_{B_s}(x_1,b_1)
  \Big\{a_i(t_a)E_e(t_a)\nonumber\\
  &&\times\Big[\phi_3^A(x_3)+r_3(2+x_3)\phi_3^P(x_3)-r_3x_3\phi_3^T(x_3)\Big]
  h_e(x_1,x_3,b_1,b_3)\nonumber\\
  &&\;\;\;+2
  r_3\phi_3^P(x_3)a_i(t^\prime_a)E_e(t_a^\prime)h_e(x_3,x_1,b_3,b_1)\Big\},\label{ppefsp}
  \end{eqnarray}
\end{itemize}
where the color factor ${C_F}=4/3$ and $a_i$ represents the corresponding Wilson coefficients from differen decay channels.
$r_i=\frac{m_{0i}}{m_{B_s}}$, where $m_{0i}$ refers to the chiral scale parameter.
\begin{itemize}
\item $(V-A)(V-A)$ operators:
\begin{eqnarray} M_{B_s\to M_3}^{LL}(a_i)&=&32\pi
C_FM_{B_s}^4/\sqrt{6}\int^1_0dx_1dx_2dx_3\int^\infty_0b_1db_1b_2db_2
\phi_{B_s}(x_1,b_1)\phi_2^A(x_2)
\nonumber\\
&&\times
\Big\{\Big[(1-x_2)\phi_3^A(x_3)-r_3x_3(\phi_3^P(x_3)-\phi_3^T(x_3))\Big]
a_i(t_b)E_e^\prime(t_b)\nonumber\\
&&~\times h_n(x_1,1-x_2,x_3,b_1,b_2)+h_n(x_1,x_2,x_3,b_1,b_2)\nonumber\\
 &&\;\;\times\Big[-(x_2+x_3)\phi_3^A(x_3)+r_3x_3(\phi_3^P(x_3)+\phi_3^T(x_3))\Big]
 a_i(t_b^\prime) E_e^\prime(t_b^\prime)\Big\},\label{ppenll}
 \end{eqnarray}

\item $(V-A)(V+A)$ operators:
\begin{eqnarray} M_{B_s\to M_3}^{LR}( a_i)&=&32\pi C_FM_{B_s}^4r_2/\sqrt{6}
      \int^1_0dx_1dx_2dx_3\int^\infty_0b_1db_1b_2db_2\phi_{B_s}(x_1,b_1)\nonumber\\
     &&\times \Big\{h_n(x_1,1-x_2,x_3,b_1,b_2)\Big[(1-x_2)\phi_3^A(x_3)
     \left(\phi_2^P(x_2)+\phi_2^T(x_2)\right)\nonumber\\
     &&\;\;+r_3x_3\left(\phi_2^P(x_2)-\phi_2^T(x_2)\right)
     \left(\phi_3^P(x_3)+\phi_3^T(x_3)\right)\nonumber\\
     &&\;\;+(1-x_2)r_3\left(\phi_2^P(x_2)+\phi_2^T(x_2)\right)\left(\phi_3^P(x_3)
      -\phi_3^T(x_3)\right)\Big]a_i(t_b)
         E_e^\prime(t_b) \nonumber\\
       &&\;\;-h_n(x_1,x_2,x_3,b_1,b_2)\Big[x_2\phi_3^A(x_3)(\phi_2^P(x_2)-\phi_2^T(x_2))\nonumber\\
       &&\;\;+r_3x_2(\phi_2^P(x_2)-\phi_2^T(x_2))(\phi_3^P(x_3)-\phi_3^T(x_3))\nonumber\\
      &&\;\;+r_3x_3(\phi_2^P(x_2)+\phi_2^T(x_2))(\phi_3^P(x_3)+\phi_3^T(x_3))\Big]a_i(t^\prime_b)
       E_e^\prime(t_b^\prime)\Big\},\label{ppenlr}
 \end{eqnarray}

\item $(S-P)(S+P)$ operators:
\begin{eqnarray} M^{SP}_{B_s\to M_3}( a_i) &=&32\pi C_F
M_{B_s}^4/\sqrt{6}\int^1_0dx_1dx_2dx_3\int^\infty_0b_1db_1b_2db_2
\phi_{B_s}(x_1,b_1)\phi_2^A(x_2)
\nonumber\\
&&\times\Big\{
\Big[(x_2-x_3-1)\phi_3^A(x_3)+r_3x_3(\phi_3^P(x_3)+\phi_3^T(x_3))\Big]\nonumber\\
&&\times
a_i(t_b)E_e^\prime(t_b)h_n(x_1,1-x_2,x_3,b_1,b_2)+a_i(t_b^\prime)
E^\prime_e(t_b^\prime)\nonumber\\
&&\times
 \Big[x_2\phi_3^A(x_3)+r_3x_3(\phi_3^T(x_3)-\phi_3^P(x_3))\Big]h_n(x_1,x_2,x_3,b_1,b_2)\Big\}.
 \label{ppensp}
\end{eqnarray}
\end{itemize}

The functions are related with the annihilation type process, whose contributions are:
\begin{itemize}
\item $(V-A)(V-A)$ operators:
\begin{eqnarray}
f_{B_s} F_{ann}^{LL}( a_i)&=&8\pi
C_FM_{B_s}^4f_{B_s}\int^1_0dx_2dx_3\int^\infty_0b_2db_2b_3db_3\Big\{a_i(t_c)
E_a(t_c)
\nonumber\\
&&
\times\Big[(x_3-1)\phi_2^A(x_2)\phi_3^A(x_3)-4r_2r_3\phi_2^P(x_2)\phi_3^P(x_3)\nonumber
\\
&&+2r_2r_3x_3\phi_2^P(x_2)(\phi_3^P(x_3)-\phi_3^T(x_3))\Big]h_a(x_2,1-x_3,b_2,b_3)\nonumber
\\
&&+\Big[x_2\phi_2^A(x_2)
\phi_3^A(x_3)+2r_2r_3(\phi_2^P(x_2)-\phi_2^T(x_2))\phi_3^P(x_3)\nonumber\\
&&+2r_2r_3x_2(\phi_2^P(x_2)+\phi_2^T(x_2))\phi_3^P(x_3)\Big]
a_i(t_c^\prime)
E_a(t_c^\prime)h_a(1-x_3,x_2,b_3,b_2)\Big\}.\label{ppafll}
 \end{eqnarray}

 \item
$(V-A)(V+A)$ operators:
\begin{eqnarray}
F_{ann}^{LR}( a_i)=F_{ann}^{LL}(a_i),\label{ppaflr}
\end{eqnarray}

 \item $(S-P)(S+P)$ operators:
 \begin{eqnarray}
 f_{B_s} F_{ann}^{SP}(a_i)&=&16\pi
  C_FM_{B_s}^4f_{B_s}\int^1_0dx_2dx_3\int^\infty_0b_2db_2b_3db_3
  \Big\{\Big[2r_2\phi_2^P(x_2)\phi_3^A(x_3)\nonumber\\
  &&\;\;+(1-x_3)r_3\phi_2^A(x_2)(\phi_3^P(x_3)
  +\phi_3^T(x_3))\Big]
 a_i(t_c) E_a(t_c)h_a(x_2,1-x_3,b_2,b_3)\nonumber\\
  &&\;\;+\Big[2r_3\phi_2^A(x_2)\phi_3^P(x_3)+r_2x_2(\phi_2^P(x_2)-\phi_2^T(x_2))\phi_3^A(x_3)
  \Big]\nonumber\\
  &&\;\;\times
  a_i(t_c^\prime)E_a(t_c^\prime)h_a(1-x_3,x_2,b_3,b_2)\Big\}.\label{ppafsp}
\end{eqnarray}
\end{itemize}

\begin{itemize}
\item $(V-A)(V-A)$ operators:
\begin{eqnarray}
 M_{ann}^{LL}( a_i)&=&32\pi C_FM_{B_s}^4/\sqrt
 {6}\int^1_0dx_1dx_2dx_3\int^\infty_0b_1db_2b_2db_2\phi_{B_s}(x_1,b_1)\nonumber\\
 &&\times \Big\{h_{na}(x_1,x_2,x_3,b_1,b_2)\Big[-x_2\phi_2^A(x_2)\phi_3^A(x_3)-4r_2r_3
 \phi_2^P(x_2)\phi_3^P(x_3)\nonumber\\
 &&\;\;\;+r_2r_3(1-x_2)(\phi_2^P(x_2)+\phi_2^T(x_2))(\phi_3^P(x_3)-\phi_3^T(x_3))
 \nonumber\\
 &&\;\;+r_2r_3x_3(\phi_2^P(x_2)-\phi_2^T(x_2))(\phi_3^P(x_3)+\phi_3^T(x_3))\Big]a_i(t_d)
 E_a^\prime(t_d)\nonumber\\
 &&\;\;+h_{na}^\prime(x_1,x_2,x_3,b_1,b_2)\Big[(1-x_3)\phi_2^A(x_2)\phi_3^A(x_3)
 \nonumber\\
 &&\;\;+(1-x_3)r_2r_3(\phi_2^P(x_2)+\phi_2^T(x_2))(\phi_3^P(x_3)-\phi_3^T(x_3))
 \nonumber\\
 &&\;\;+x_2r_2r_3(\phi_2^P(x_2)-\phi_2^T(x_2))(\phi_3^P(x_3)+\phi_3^T(x_3))\Big]
 a_i(t_d^\prime)
 E_a^\prime(t_d^\prime)\Big\},\label{ppanll}
 \end{eqnarray}

 \item $(V-A)(V+A)$ operators:
 \begin{eqnarray}
 M_{ann}^{LR}(M_2,M_3, a_i)&=&32\pi C_FM_{B_s}^4/\sqrt
 {6}\int^1_0dx_1dx_2dx_3\int^\infty b_1db_1b_2db_2\phi_{B_s}(x_1,b_1)\nonumber\\
 &&\;\;\times\Big\{h_{na}(x_1,x_2,x_3,b_1,b_2)\Big[r_2(2-x_2)(\phi_2^P(x_2)+\phi_2^T(x_2))
 \phi_3^A(x_3)\nonumber\\
 &&\;\;-r_3(1+x_3)\phi_2^A(x_2)(\phi_3^P(x_3)-\phi_3^T(x_3))\Big]a_i(t_d)E_a^\prime(t_d)
 \nonumber\\
 &&\;\;+h_{na}^\prime
 (x_1,x_2,x_3,b_1,b_2)\Big[r_2x_2\left(\phi_2^P(x_2)+\phi_2^T(x_2)\right)\phi_3^A(x_3)
 \nonumber\\
 &&\;\;+r_3(x_3-1)\phi_2^A(x_2)
 (\phi_3^P(x_3)-\phi_3^T(x_3))\Big]  a_i(t_d^\prime)E_a^\prime(t_d^\prime)
 \Big\},\label{ppanlr}
 \end{eqnarray}

 \item $(S-P)(S+P)$ operators:
 \begin{eqnarray}
 M_{ann}^{SP}( a_i)&=&32\pi C_F M_{B_s}^4/\sqrt {6}\int^1_0dx_1dx_2dx_3\int^\infty_0b_1db_1b_2db_2
 \phi_{B_s}(x_1,b_1)\nonumber\\
 &&\times \Big\{a_i(t_d)E_a^\prime(t_d)h_{na}(x_1,x_2,x_3,b_1,b_2)\Big[(x_3-1)
 \phi_2^A(x_2)\phi_3^A(x_3)\nonumber\\
 &&\;\; -4r_2r_3\phi_2^P(x_2)\phi_3^P(x_3)+r_2r_3x_3(\phi_2^P(x_2)+\phi_2^T(x_2))
 (\phi^P_3(x_3)-\phi_3^T(x_3))\nonumber\\
 &&\;\;+r_2r_3(1-x_2)(\phi_2^P(x_2)-\phi_2^T(x_2))(\phi^P_3(x_3)+\phi_3^T(x_3))\Big]
 \nonumber\\
 &&\;\;+a_i(t_d^\prime)
 E_a^\prime(t_d^\prime)h_{na}^\prime(x_1,x_2,x_3,b_1,b_2)
  \Big[x_2\phi_2^A(x_2)\phi_3^A(x_3)\nonumber
 \\
 &&\;\;+x_2r_2r_3(\phi_2^P(x_2)+\phi_2^T(x_2))
 (\phi_3^P(x_3)-\phi_3^T(x_3)))\nonumber\\
 &&\;\;+r_2r_3(1-x_3)(\phi_2^P(x_2)-\phi_2^T(x_2))(\phi_3^P(x_3)+\phi_3^T(x_3))\Big]\Big\}.
 \label{ppansp}
 \end{eqnarray}
\end{itemize}



\end{spacing}

\begin{thebibliography}{}
\bibitem{cab}N. Cabibbo, Phys. Rev. Lett. {\bf10}, 531 (1963).

\bibitem{kob}M. Kobayashi and T. Maskawa, Prog.
Theor. Phys. {\bf49}, 652 (1973).

\bibitem{AC}A. Ali and C. Greub, Phys. Rev. {\bf D57}, 2996 (1998); G. Kramer, W. F. Palmer, and H. Simma, Nucl. Phys. {\bf B428}, 77 (1994); Z. Phys. {\bf C66}, 429 (1995).



\bibitem{AG}A. Ali, G. Kramer, and C.-D. Lu, Phys. Rev. {\bf D58}, 094009 (1998); {\bf59}, 014005 (1998); Y. H. Chen, H. Y. Cheng, B. Tseng, and K. C. Yang, Phys. Rev. {\bf D60}, 094014 (1999).


\bibitem{YA}Y. Y. Keum, H.-n. Li, and A. I. Sanda, Phys. Rev. Lett. {\bf B504}, 6 (2001); Phys. Rev. {\bf D63}, 054008 (2001).
\bibitem{LKM}C.-D. Lu, K. Ukai, and M.-Z. Yang, Phys. Rev. {\bf D63}, 074009 (2001).

\bibitem{MGronau1990} M. Gronau and D. London, Phys.
Rev. Lett. {\bf 65}, 3381 (1990).

\bibitem{gar1998}S. Gardner, H.B. O'Connell, and A.W. Thomas, Phys.
Rev. Lett. {\bf 80}, 1834 (1998).


\bibitem{gang4}Gang L$\ddot{u}$, Ye Lu, Sheng-Tao Li and Yu-Ting Wang,
Eur. Phys. J. {\bf C77}, 518 (2017).


\bibitem{garden1999M} S. Gardner, Phys. Rev. {\bf D59}, 077502 (1999).

\bibitem{Gronau2005J} M. Gronau and J. Zupan, Phys. Rev. {\bf D71}, 074017 (2005).

\bibitem{Kroll2005PP} P. Kroll, Modern Physics Letters {\bf A20}, 2667 (2005).

\bibitem{AOsip2016} A. A. Osipov, B. Hiller and A. H. Blin, Phys. Rev. {\bf D93}, 116005 (2016).




\bibitem{buras} G. Buchalla, A.~J. Buras,
M.~E. Lautenbacher, Rev. Mod. Phys. {\bf68}, 1125 (1996).
\bibitem{YH}Y.H. Chen, H.Y. Cheng, B.Tseng, K.C. Yang, Phys. Rev. D {\bf60}, 094014 (1999).

\bibitem{AMLi2007}Ahmed Ali, Gustav Kramer, Ying Li, Cai-Dian Lu, Yue-Long Shen, Wei Wang, Phys. Rev. {\bf D76}, 074018 (2007).

\bibitem{tpbm} Th. Feldmann, P. Kroll, and B. Stech, Phys. Rev. {\bf D58}, 114006 (1998);  Phys. Rev. Lett. {\bf B449}, 339 (1999).






\bibitem{wangz2014} J.-J. Wang, D.-T. Lin, W. Sun, Z.-J. Ji, S. Cheng, and Z.-J. Xiao, Phys. Rev. {\bf D89}, 074046 (2014).


\bibitem{hnl2003}H.-n. Li, Prog. Part. Nucl. Phys, {\bf D51}, 85 (2003).


\bibitem{wol}L. Wolfenstein, Phys. Rev. Lett. {\bf 51}, 1945 (1983); Phys. Rev. Lett. {\bf13}, 562 (1964).
\bibitem{PDG2018}M. Tanabashi et al. (Particle Data Group), Phys. Rev. {\bf D89}, 030001 (2018).


\bibitem{Aaij:2014jba} R. Aaij et al. (LHCb), Int. J. Mod. Phys. {\bf A30}, 1530022 (2015).


\bibitem{Szumlak:2017und} T. Szumlak, J. Phys. Conf. Ser. {\bf 898}, 032051 (2017).

\bibitem{Whitehead:2017vob} M. P. Whitehead (LHCb), PoS {\bf EPS-HEP2017}, 528 (2017).



\bibitem{Chen:2017khz} X. Chen (ATLAS, CMS), in Proceedings, International Workshop on
Future Linear Colliders 2016 (LCWS2016): Morioka, Iwate, Japan, December 05-09, 2016 (2017), 1703.07689.



\bibitem{Schopper:2006he} A. Schopper, eConf {\bf C060409}, 042 (2006).
\bibitem{LHC2017} R. Aaij et al. (LHCb), Phys. Rev. {\bf D95}, 012006 (2017).
\bibitem{L3} H.-n. Li, Phys. Rev. {\bf D66}, 094010 (2002).
\bibitem{L4} H.-n. Li and K. Ukai, Phys. Lett. {\bf B555}, 197 (2003).


\end{thebibliography}
\end{document}